# Lattice Anharmonicity in $BiS_2$-Based Layered Superconductor $RE(O,F)BiS_2$ (RE = La, Ce, Pr, Nd)

Fysol Ibna Abbas[1,2], Kazuhisa Hoshi[1], Aichi Yamashita[1], Yuki Nakahira[1], Yosuke Goto[1], Akira Miura[3], Chikako Moriyoshi[4], Yoshihiro Kuroiwa[4], Kensei Terashima[5], Ryo Matsumoto[5,6], Yoshihiko Takano[5], and Yoshikazu Mizuguchi[1]*

[1] *Department of Physics, Tokyo Metropolitan University, 1-1, Minami-osawa, Hachioji 192-0397, Japan*
[2] *Department of Electrical & Electronic Engineering, City University, Khagan, Birulia, Savar, Dhaka-1216, Bangladesh*
[3] *Faculty of Engineering, Hokkaido University, Kita 13 Nishi 8, Sapporo 060-8628, Japan*
[4] *Graduate School of Advanced Science and Engineering, Hiroshima University, Higashihiroshima, Hiroshima, 739-8526, Japan*
[5] *International Center for Materials Nanoarchitectonics (MANA), National Institute for Materials Science, Tsukuba, Ibaraki 305-0047, Japan*
[6] *International Center for Young Scientists (ICYS), National Institute for Materials Science, Tsukuba, Ibaraki 305-0047, Japan*
*6*

**Abstract**

We studied Grüneisen parameter ($\gamma_G$) to investigate lattice anharmonicity in a layered $BiS_2$-based superconductor system $REO_{1-x}F_xBiS_2$ (RE = La, Ce, Pr, Nd), where in-plane chemical pressure was tuned by substituting the RE elements. With increasing in-plane chemical pressure, bulk modulus remarkably increases, and a high $\gamma_G$ is observed for RE = Nd. The dependence of $\gamma_G$ on in-plane chemical pressure exhibits a good correlation with $T_c$, and a higher $T_c$ is achieved when $\gamma_G$ is large for RE = Nd. In addition, $\gamma_G$ shows a slight decrease by a decrease of F concentration ($x$) in $REO_{1-x}F_xBiS_2$. Our results show that the anharmonic vibration of Bi along the $c$-axis is present in $REO_{1-x}F_xBiS_2$, and the enhancement of the anharmonicity is positively linked to superconducting $T_c$ and pairing mechanisms.

## 1. Introduction

The presence of lattice anharmonicity has been one of the important issues in superconductors with a high transition temperature ($T_c$) and exotic pairing mechanisms. For example, anharmonicity and structural instability is essential parameter for $T_c$ in A15-type superconductors [1] and rattling materials [2,3]. In addition, anharmonicity has been observed in $MgB_2$ [4] and superhydrides (superconductors with hydrogen-rich compositions) [5]. Although anharmonic vibration is positively linked to superconductivity in [2,3,6], a theoretical study on a superhydride suggested that giant anharmonicity suppresses $T_c$ in the material [7]. Therefore, control of anharmonicity is one of key factors for designing new superconductors and improving their superconducting properties. In this study, we investigate the relationship between superconductivity and anharmonicity in a $BiS_2$-based layered superconductor via the estimation of Grüneisen parameter ($\gamma_G$).

The $BiS_2$-based superconductors, which are a layered superconductor family having an alternate stacking structure of blocking layer and conducting layer ($BiS_2$ layer), were discovered in 2012, and many analogous $BiS_2$-based superconductors have been synthesized [8–19]. $BiS_2$ layers are essential to superconductivity [20–22], and superconductivity is induced after carrier doping and optimization of crystal structure [17,23,24]. In particular, the importance of in-plane chemical pressure effect [25], which suppresses local structural disorder (local distortion) in $BiS_2$ layers [25–30], has been found. One of the origins of the local disorder in the $BiS_2$ layers is the presence of Bi lone pairs [27]. In addition, structural instability has been predicted theoretically [31]. Hence, the relationship between superconductivity and structural instability would be a kye factor, but the point has not been experimentally studied. On the mechanisms of superconductivity, recent experimental and theoretical works indicated the possibility of unconventional superconductivity with an anisotropic superconducting gap structure [32–36]. Furthermore, by focusing on the suppression of the local disorder and the improvement of superconducting properties by in-plane chemical pressure [25], we have revealed some evidences of unconventional pairing in La(O,F)Bi(S,Se)$_2$ and $Bi_4O_4S_3$ where in-plane chemical pressure is high enough for the suppression of local disorder: the absence of isotope effects, nematic superconductivity, and high upper critical fields [38–42].

Motivated by the background knowledges on $BiS_2$-based superconductivity, we here focus on the relationship between superconductivity and lattice anharmonicity in $REO_{1-x}F_xBiS_2$ (RE = La, Ce, Pr, Nd). In a related LaOBi(S,Se)$_2$ system, which shows high thermoelectric properties [43,44], evolution of anharmonic vibration in $BiS_2$ layers has been detected by inelastic neutron scattering (INS) [45]. In the system, low thermal conductivity was explained by the enhanced anharmonicity. In addition, we recently analyzed anharmonicity in $LaOBi(S,Se)_2$ by estimating $\gamma_G$ and reproduced the trend found in the INS [46]. Since anharmonicity is enhanced by in-plane chemical pressure in $LaOBi(S,Se)_2$ by Se substitution, we investigate the dependences of $\gamma_G$ on in-plane chemical pressure tuned by RE ionic radii in $REO_{1-x}F_xBiS_2$ in this paper. As a main conclusion, we show that anharmonicity is enhanced by in-plane chemical pressure and exhibits a positive correlation with $T_c$ in $REO_{1-x}F_xBiS_2$.

## 2. Experimental details

*2.1 methods*

Polycrystalline samples were synthesized by solid-state reaction as described in Refs. 9 and 26. Synchrotron X-ray diffraction (SXRD) experiments were performed at the beamline BL02B2, SPring-8 at a wavelength of 0.49559 Å (proposal No. 2017B1211). The SXRD experiments were performed with a sample rotator system at room temperature; the diffraction data were collected using a high-resolution one-dimensional semiconductor detector (multiple MYTHEN system [47]) with a step size of $2\theta$ = 0.006°. The temperature of the sample was controlled by nitrogen gas. The crystal structure parameters were refined using the Rietveld method using RIETAN-FP [48]. The image of schematic models of the crystal structures was produced using VESTA [49]. Specific heat was measured by relaxation method using a Physical Property Measurement System (PPMS, Quantum Design). Longitudinal sound velocity ($v_L$) of the sample was measured using an ultrasonic thickness detector (Satotech-2000C). For sound velocity measurements, the obtained samples were densified using high-pressure synthesis method with a 180-ton cubic-anvil system with a pressure of ~1.5 GPa and an annealing temperature of ~400 °C for 15 min.

*2.2 Grüneisen parameter*

$\gamma_G$ of samples was calculated using the following formula [45,50];

$$\gamma_G = \frac{\beta_V B V_{mol}}{C_V} \qquad (1)$$

, where $\beta_V$, $B$, $V_{mol}$, and $C_V$ are volume thermal expansion coefficient, bulk modulus, molar volume, and specific heat, respectively. The parameters needed for the estimation of $\gamma_G$ are calculated as follows (formula 2–5). In the formulas, $dV/dT$, $\rho$, $v_L$, $v_S$, $v_m$, $\theta_D$, $h$, $k_B$, $n$, $N_A$, and $M$ denote temperature gradient of lattice volume, density of the material, longitudinal sound velocity, shear sound velocity, average sound velocity, Debye temperature, Plank's constant,

Boltzmann's constant, number of atoms in the molecule (formula unit), Avogadro's constant, and the molecular weight (per formula unit).

$$\beta_{\mathrm{V}} = \frac{1}{V(300\ \mathrm{K})}\frac{dV}{dT} \qquad (2)$$

$$B = \rho\left(v_{\mathrm{L}}^2 - \frac{4}{3}v_{\mathrm{S}}^2\right) \qquad (3)$$

$$\theta_D = \left(\frac{h}{k_{\mathrm{B}}}\right)\left[\frac{3n}{4\pi}\left(\frac{N_{\mathrm{A}}\rho}{M}\right)\right]^{\frac{1}{3}} v_{\mathrm{m}} \qquad (4)$$

$$v_m = \left[\frac{1}{3}\left(\frac{2}{v_{\mathrm{S}}^3} + \frac{1}{v_{\mathrm{L}}^3}\right)\right]^{-\frac{1}{3}} \qquad (5)$$

## 3. Results and Discussion

Figure 1 displays the SXRD patterns collected at $T$ = 600 K and the Rietveld refinement results. The results suggest that the high-temperature measurements were successfully performed. In the refinement, tiny impurity phases of $REF_3$ were assumed [26]. The estimated lattice volume ($V$) is plotted in Fig. 2. By formula (2) and linear fitting in Fig. 2a–2e, $\beta_{\mathrm{V}}$ is estimated as 39.6, 32.9, 38.4, 37.8, and 36.4 ($\mu K^{-1}$) for $LaO_{0.5}F_{0.5}BiS_2$, $CeO_{0.5}F_{0.5}BiS_2$, $PrO_{0.5}F_{0.5}BiS_2$, $NdO_{0.5}F_{0.5}BiS_2$, and $NdO_{0.7}F_{0.3}BiS_2$, respectively.

The temperature dependences of the isotropic displacement parameters ($U_{\mathrm{iso}}$) are plotted in Fig. 2g–2j. As mentioned in introduction, $LaO_{0.5}F_{0.5}BiS_2$ exhibits huge $U_{\mathrm{iso}}$ at the S1 site (see Fig. 2f), which is due to low in-plane chemical pressure [25,26]. The temperature dependence of $U_{\mathrm{iso}}$ for S1 is not remarkable, and the $U_{\mathrm{iso}}$ expected at $T$ = 0 K is clearly large, which indicate the presence of static disorder at the S1 site. With increasing in-plane chemical pressure (by replacing La by Ce or Pr), $U_{\mathrm{iso}}$ at the S1 site is slightly suppressed. For RE = Nd, $U_{\mathrm{iso}}$ at the S1 site is clearly decreased and becomes lower than that at the Bi site. Furthermore, the $U_{\mathrm{iso}}$ at the S1 site is expected to approach zero at $T$ = 0 K from the temperature dependence as shown in Fig. 2j, suggesting the suppression of in-plane static disorder in the superconducting sample with RE = Nd.

In Fig. 3a–3d, the $T^2$ dependence of $C/T$ are displayed, and $\theta_{\mathrm{D}}$ is estimated as 195, 197, 222, and 198 K for $LaO_{0.5}F_{0.5}BiS_2$, $CeO_{0.5}F_{0.5}BiS_2$, $PrO_{0.5}F_{0.5}BiS_2$, and $NdO_{0.7}F_{0.3}BiS_2$, respectively. The slightly larger $\theta_{\mathrm{D}}$ for RE = Pr is consistent with a previous work [13].$\theta_{\mathrm{D}}$ = 189 K for $NdO_{0.5}F_{0.5}BiS_2$ was calculated from a previous work [51].

The sound velocity ($v_{\mathrm{L}}$) was measured and corrected using packing factor as described in Ref. 46. The estimated sound velocity parameters are listed in Table I. To estimate $v_{\mathrm{S}}$ and $v_{\mathrm{m}}$, formulas (3–5) were used. In addition, bulk modulus ($B$) was calculated. $B$ remarkably increases with decreasing lattice constant (increasing in-plane chemical pressure) from La to Nd as shown in Fig. 4. $B$ is given by $B = -V_0(\mathrm{d}P/\mathrm{d}V)$, where $V_0$ is volume at initial state, and $\mathrm{d}P/\mathrm{d}V$ corresponds to the change in internal pressure by the change in volume. Therefore, $B$ shows how the material is resistant to compression. The results shown in Fig. 4 clearly indicate that in-plane chemical pressure modified the $REOBiS_2$-type structure, and the huge changes in structural characteristics should affect lattice vibration characters in the $REO_{1-x}F_xBiS_2$ system.

We estimated $\gamma_G$ for the $REO_{1-x}F_xBiS_2$ samples to examine the evolution of anharmonicity by in-plane chemical pressure and the relationship with superconductivity. In addition, the influence of carrier doping amount is also discussed for RE = Pr and Nd. Figure 5(a) displays the $T_c$ for $REO_{0.5}F_{0.5}BiS_2$ ($x = 0.5$) plotted as a function of $RE^{3+}$ ionic radii [26]. In Fig. 5(b), the estimated $\gamma_G$ is plotted. For $\gamma_G$ ($x = 0.5$), the trend exhibits a correlation to that for $T_c$; $T_c$ increases with increasing $\gamma_G$. On the basis of previous studies on $LaOBiS_{2-x}Se_x$ by INS [45] and $\gamma_G$ [46], we assume that the increase in $\gamma_G$ in $REO_{1-x}F_xBiS_2$ systems is also originating from the modified vibration of Bi along the $c$-axis [45].

On the effect of carrier concentration ($x$) on $\gamma_G$ in $REO_{1-x}F_xBiS_2$, we observed a trend that $\gamma_G$ for $x = 0.3$ is slightly smaller than that for $x = 0.5$ [Fig. 5(b)]. This difference would be due to the in-plane structural instability. As described in introduction, the in-plane local disorder in doped (F-substituted) $REO_{1-x}F_xBiS_2$ is suppressed by in-plane chemical pressure. However, the local disorder is remarkable in non-doped or low-carrier phases as suggested in Ref. 52. As a fact, the in-plane local disorder in $REO_{1-x}F_xBiS_2$ is suppressed by both F substitution and in-plane chemical pressure, and the anharmonic vibration is enhanced in less-disordered phases, namely in $NdO_{0.5}F_{0.5}BiS_2$.

We conclude that, with increasing in-plane chemical pressure, the anharmonic nature of the Bi vibration is enhanced, and $T_c$ increases in the system. If the assumption is correct, we expect a higher $T_c$ when the anharmonicity is further enhanced by applying a higher in-plane chemical pressure; for example, uniaxial strain on single crystals or strain effects in thin films would achieve such a situation. To clarify the nature of lattice vibration in $REO_{1-x}F_xBiS_2$ in detail, INS experiments are desired. In addition, further understanding of electronic states will be helpful for understanding the mechanisms of superconductivity in $BiS_2$-based systems because the enhancement of anharmonic nature generally enhances electron-phonon coupling [2]. Investigation on lattice anharmonicity in $BiS_2$-based systems will open new pathway to develop exotic superconductors where anharmonic vibration plays an essential role on pairing.

## 4. Conclusion

We have investigated Grüneisen parameter ($\gamma_G$) of $REO_{1-x}F_xBiS_2$ with different RE ions and carrier concentrations through the measurements of temperature-dependent XRD, sound velocity, and low-temperature specific heat ($\theta_D$). With increasing in-plane chemical pressure, bulk modulus remarkably increases, and a high $\gamma_G$ is observed for RE = Nd. The dependence of $\gamma_G$ on in-plane chemical pressure exhibits a good correlation with $T_c$, and a higher $T_c$ is achieved when $\gamma_G$ is large for RE = Nd. In addition, $\gamma_G$ shows a slight decrease by a decrease of F concentration ($x$) in $REO_{1-x}F_xBiS_2$. We propose that the anharmonic vibration of Bi along the $c$-axis observed in previous inelastic neutron scattering for thermoelectric $LaOBiS_{2-x}Se_x$ is present in $REO_{1-x}F_xBiS_2$, and the enhancement of the anharmonicity is positively linked to superconducting $T_c$ and pairing mechanisms.

### Acknowledgment

This work was partially funded by Grant-in-Aid for Scientific Research (KAKENHI) (No. 18KK0076) and Tokyo Metropolitan Government Advanced Research (No. H31-1).

*E-mail: mizugu@tmu.ac.jp

1) L. R. Testardi, Phys. Rev. B 5, 4342 (1972).
2) Z. Hiroi, J. Yamaura, and K. Hattori, J. Phys. Soc. Jpn. **81**, 011012 (2012).
3) K. Oshiba and T. Hotta, J. Phys. Soc. Jpn. **80**, 094712 (2011).
4) A. Y. Liu, I. I. Mazin, and J. Kortus, Phys. Rev. Lett. **87**, 087005 (2001).
5) I. Errea, Matteo Calandra, C. J. Pickard, J. Nelson, R. J. Needs, Y. Li, H. Liu, Y. Zhang, Y. Ma, and F. Mauri, Phys. Rev. Lett. **114**, 157004 (2015).
6) T. Isono, D. Iguchi, T. Matsubara, Y. Machida, B. Salce, J. Flouquet, H. Ogusu, J. I. Yamaura, Z. Hiroi, and K. Izawa, J. Phys. Soc. Jpn. **82**, 114708 (2013).
7) B. Rousseau and A. Bergara, Phys. Rev. B **82**, 104504 (2010).
8) Y. Mizuguchi, H. Fujihisa, Y. Gotoh, K. Suzuki, H. Usui, K. Kuroki, S. Demura, Y. Takano, H. Izawa, and O. Miura, Phys. Rev. B **86**, 220510 (2012).
9) Y. Mizuguchi, S. Demura, K. Deguchi, Y. Takano, H. Fujihisa, Y. Gotoh, H. Izawa, and O. Miura, J. Phys. Soc. Jpn. **81**, 114725 (2012).

10) S. Demura, Y. Mizuguchi, K. Deguchi, H. Okazaki, H. Hara, T. Watanabe, S. J. Denholme, M. Fujioka, T. Ozaki, H. Fujihisa, Y. Gotoh, O. Miura, T. Yamaguchi, H. Takeya, and Y. Takano, J. Phys. Soc. Jpn. **82**, 033708 (2013).
11) J. Xing, S. Li, X. Ding, H. Yang, and H. H. Wen, Phys. Rev. B **86**, 214518 (2012).
12) S. Demura, K. Deguchi, Y. Mizuguchi, K. Sato, R. Honjyo, A. Yamashita, T. Yamaki, H. Hara, T. Watanabe, S. J. Denholme, M. Fujioka, H. Okazaki, T. Ozaki, O. Miura, T. Yamaguchi, H. Takeya, and Y. Takano, J. Phys. Soc. Jpn. **84**, 024709 (2015).
13) R. Jha, A. Kumar, S. K. Singh, V. P. S. Awana, J. Supercond. Nov. Magn. **26**, 499 (2012).
14) D. Yazici, K. Huang, B. D. White, A. H. Chang, A. J. Friedman, and M. B. Maple, Philos. Mag. **93**, 673 (2012).
15) X. Lin, X. Ni, B. Chen, X. Xu, X. Yang, J. Dai, Y. Li, X. Yang, Y. Luo, Q. Tao, G. Cao, and Z. Xu, Phys. Rev. B **87**, 020504 (2013).
16) H. F. Zhai, Z. T. Tang, H. Jiang, K. Xu, K. Zhang, P. Zhang, J. K. Bao, Y. L. Sun, W. H. Jiao, I. Nowik, I. Felner, Y. K. Li, X. F. Xu, Q. Tao, C. M. Feng, Z. A. Xu, and G. H. Cao, Phys. Rev. B **90**, 064518 (2014).
17) Y. Mizuguchi, J. Phys. Soc. Jpn. **88**, 041001 (2019).
18) H. F. Zhai, P. Zhang, and G. H. Cao, J. Phys. Soc. Jpn. **88**, 041003 (2019).
19) S. Iwasaki, Y. Kawai, S. Takahashi, T. Suda, Y. Wang, Y. Koshino, F. Ogura, Y. Shibayama, T. Kurosawa, M. Oda, M. Ido, and N. Momono, J. Phys. Soc. Jpn. **88**, 041005 (2019).
20) H. Usui, K. Suzuki, and K. Kuroki, Phys. Rev. B **86**, 220501 (2012).
21) H. Usui and K. Kuroki, Nov. Supercond. Mater. **1**, 50 (2015).
22) K. Terashima, J. Sonoyama, T. Wakita, M. Sunagawa, K. Ono, H. Kumigashira, T. Muro, M. Nagao, S. Watauchi, I. Tanaka, H. Okazaki, Y. Takano, O. Miura, Y. Mizuguchi, H. Usui, K. Suzuki, K. Kuroki, Y. Muraoka, and T. Yokoya, Phys. Rev. B **90**, 220512 (2014).
23) Y. Mizuguchi, T. Hiroi, J. Kajitani, H. Takatsu, H. Kadowaki, and O. Miura, J. Phys. Soc. Jpn. **83**, 053704 (2014).
24) J. Kajitani, T. Hiroi, A. Omachi, O. Miura, and Y. Mizuguchi, J. Phys. Soc. Jpn. **84**, 044712 (2015).
25) Y. Mizuguchi, A. Miura, J. Kajitani, T. Hiroi, O. Miura, K. Tadanaga, N. Kumada, E. Magome, C. Moriyoshi, and Y. Kuroiwa, Sci. Rep. **5**, 14968 (2015).
26) Y. Mizuguchi, K. Hoshi, Y. Goto, A. Miura, K. Tadanaga, C. Moriyoshi, and Y. Kuroiwa, J. Phys. Soc. Jpn. **87**, 023704 (2018).
27) Y. Mizuguchi, E. Paris, T. Sugimoto, A. Iadecola, J. Kajitani, O. Miura, T. Mizokawa, and

N. L. Saini, Phys. Chem. Chem. Phys. **17**, 22090 (2015).
28) E. Paris, Y. Mizuguchi, M. Y. Hacisalihoglu, T. Hiroi, B. Joseph, G. Aquilanti, O. Miura, T. Mizokawa, and N. L. Saini, J. Phys.: Condens. Matter 29, 145603 (2017)
29) A. Athauda, J. Yang, S. Lee, Y. Mizuguchi, K. Deguchi, Y. Takano, O. Miura, and D. Louca, Phys. Rev. B **91**, 144112 (2014)
30) A. Athauda and D. Louca, J. Phys. Soc. Jpn. **88**, 041004 (2019).
31) T. Yildirim, Phys. Rev. B **87**, 020506 (2013).
32) C. Morice, R. Akashi, T. Koretsune, S. S. Saxena, and R. Arita, Phys. Rev. B **95**, 180505 (2017).
33) K. Suzuki, H. Usui, K. Kuroki, T. Nomoto, K. Hattori, and H. Ikeda, J. Phys. Soc. Jpn. **88**, 041008 (2019).
34) J. Liu, D. Fang, Z. Wang, J. Xing, Z. Du, X. Zhu, H. Yang, and H. H. Wen, EPL **106**, 67002 (2014).
35) Y. Ota, K. Okazaki, H. Q. Yamamoto, T. Yamamoto, S. Watanabe, C. Chen, M. Nagao, S. Watauchi, I. Tanaka, Y. Takano, and S. Shin, Phys. Rev. Lett. **118**, 167002 (2017).
36) Y. C. Chan, K. Y. Yip, Y. W. Cheung, Y. T. Chan, Q. Niu, J. Kajitani, R. Higashinaka, T. D. Matsuda, Y. Yanase, Y. Aoki, K. T. Lai, and S. K. Goh, Phys. Rev. B **97**, 104509 (2018).
37) K. Hoshi and Y. Mizuguchi, J. Phys.: Condens. Matter 33, 473001 (2021).
38) K. Hoshi, Y. Goto, and Y. Mizuguchi, Phys. Rev. B **97**, 094509 (2018).
39) R. Jha and Y. Mizuguchi, Appl. Phys. Express **13**, 093001 (2020).
40) K. Hoshi, M. Kimata, Y. Goto, T. D. Matsuda, and Y. Mizuguchi, J. Phys. Soc. Jpn. **88**, 033704 (2019).
41) K. Hoshi, M. Kimata, Y. Goto, A. Miura, C. Moriyoshi, Y. Kuroiwa, M. Nagao, and Y. Mizuguchi, J. Phys. Commun. **4**, 095028 (2020).
42) K. Hoshi, R. Kurihara, Y. Goto, M. Tokunaga, and Y. Mizuguchi, Sci. Rep. **12**, 288 (2022).
43) A. Nishida, O. Miura, C. H. Lee, and Y. Mizuguchi, Appl. Phys. Express **8**, 111801 (2015).
44) Y. Mizuguchi, A. Nishida, A. Omachi, and O. Miura, Cogent Phys. **3**, 1156281(1-14) (2016).
45) C. H. Lee, A. Nishida, T. Hasegawa, H. Nishiate, H. Kunioka, S. Ohira-Kawamura, M. Nakamura, K. Nakajima, and Y. Mizuguchi, Appl. Phys. Lett. **112**, 023903 (2018).
46) F. I. Abbas, A. Yamashita, K. Hoshi, R. Kiyama, Md. R. Kasem, Y. Goto, and Y. Mizuguchi, Appl. Phys. Express **14**, 071002 (2021).
47) S. Kawaguchi, M. Takemoto, K. Osaka, E. Nishibori, C. Moriyoshi, Y. Kubota, Y. Kuroiwa, and K. Sugimoto, Rev. Sci. Instrum. **88**, 085111 (2017).

48) F. Izumi and K. Momma, Solid State Phenom. **130**, 15 (2007).
49) K. Momma and F. Izumi, J. Appl. Crystallogr. **41**, 653 (2008).
50) F. I. Abbas, Y. Nakahira, A. Yamashita, Md. R. Kasem, M. Yoshida, Y. Goto, A. Miura, K. Terashima, R. Matsumoto, Y. Takano, C. Moriyoshi, and Y. Mizuguchi, arXiv:2202.12516.
51) R. Jha, A. Kumar, S. K. Singh, and V. P. S. Awana, J. Appl. Phys. **115**, 056102 (2013).
52) N. Hirayama, M. Ochi, and K. Kuroki, Phys. Rev. B **100**, 125201 (2019).

Table I. Parameters used for the estimation of $\gamma_G$.

| RE | La | Ce | Pr | Nd | Nd |
|---|---|---|---|---|---|
| F contents ($x$) | 0.5 | 0.5 | 0.5 | 0.5 | 0.3 |
| Relative density (%) | 100 | 100 | 97.8 | 96.0 | 94.9 |
| $\beta_V$ ($\mu K^{-1}$) | 39.6 | 32.9 | 38.4 | 37.8 | 36.4 |
| $v_L$ (exp) (m/s) | 3330 | 3320 | 3450 | 3470 | 3450 |
| $v_L$ (ideal) (m/s) | 3330 | 3320 | 3527 | 3616 | 3635 |
| $v_S$ (ideal) (m/s) | 1689 | 1687 | 1875 | 1567 | 1640 |
| $v_m$ (ideal) (m/s) | 1893 | 1890 | 2096 | 1770 | 1850 |
| $\theta_D$ (K) | 195 | 197 | 222 | 189 [51] | 198 |
| $B$ (GPa) | 43.2 | 44.3 | 49.8 | 65.6 | 64.5 |
| $C_V$ (J/K mol) | 124.6 | 124.6 | 124.6 | 124.6 | 124.6 |
| $\gamma_G$ | 0.99 | 0.82 | 1.03 | 1.29 | 1.22 |

## Figures

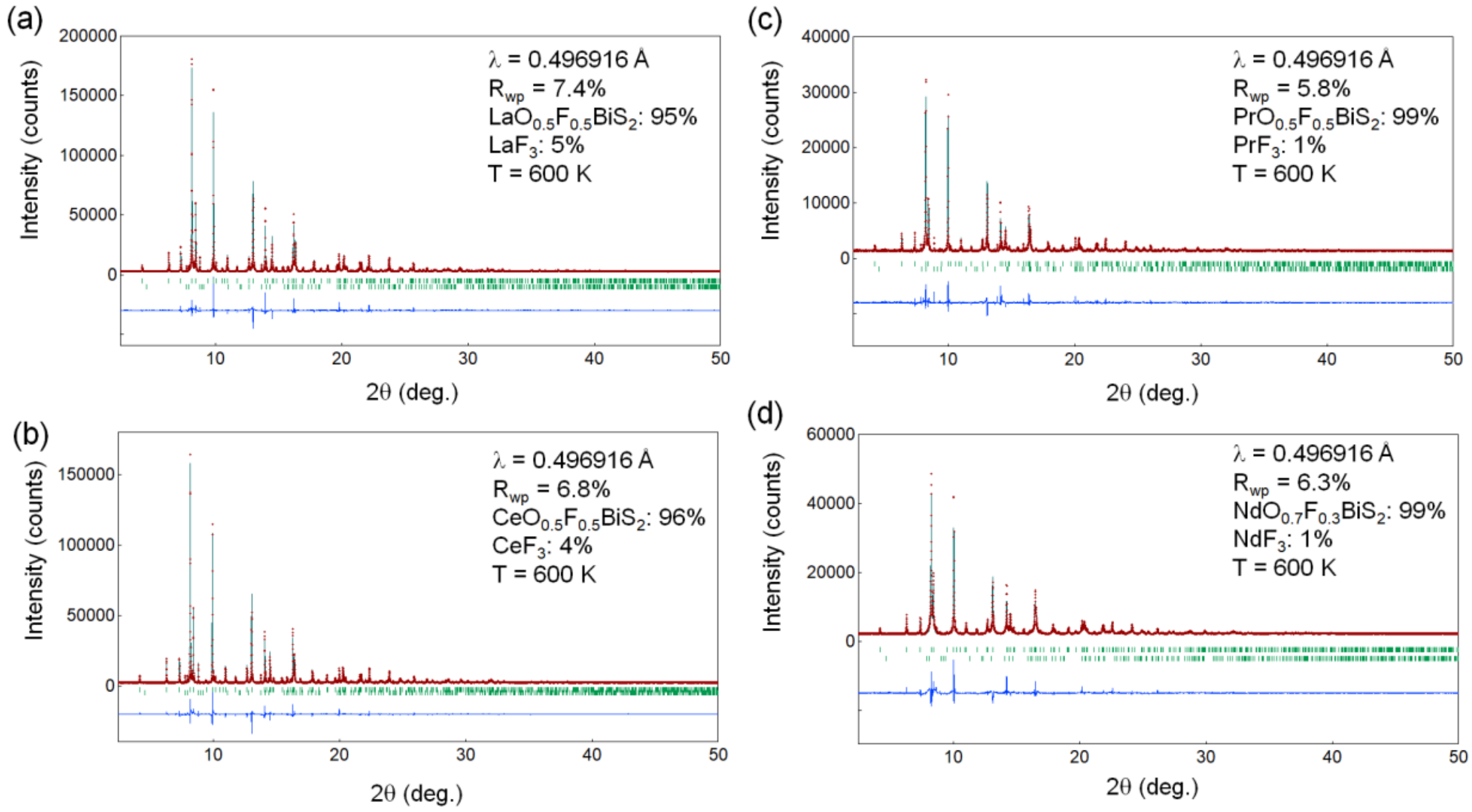

Fig. 1. SXRD patterns ($T$ = 600 K) and Rietveld refinement results for (a) $LaO_{0.5}F_{0.5}BiS_2$, (b) $CeO_{0.5}F_{0.5}BiS_2$, (c) $PrO_{0.5}F_{0.5}BiS_2$, and (d) $NdO_{0.7}F_{0.3}BiS_2$.

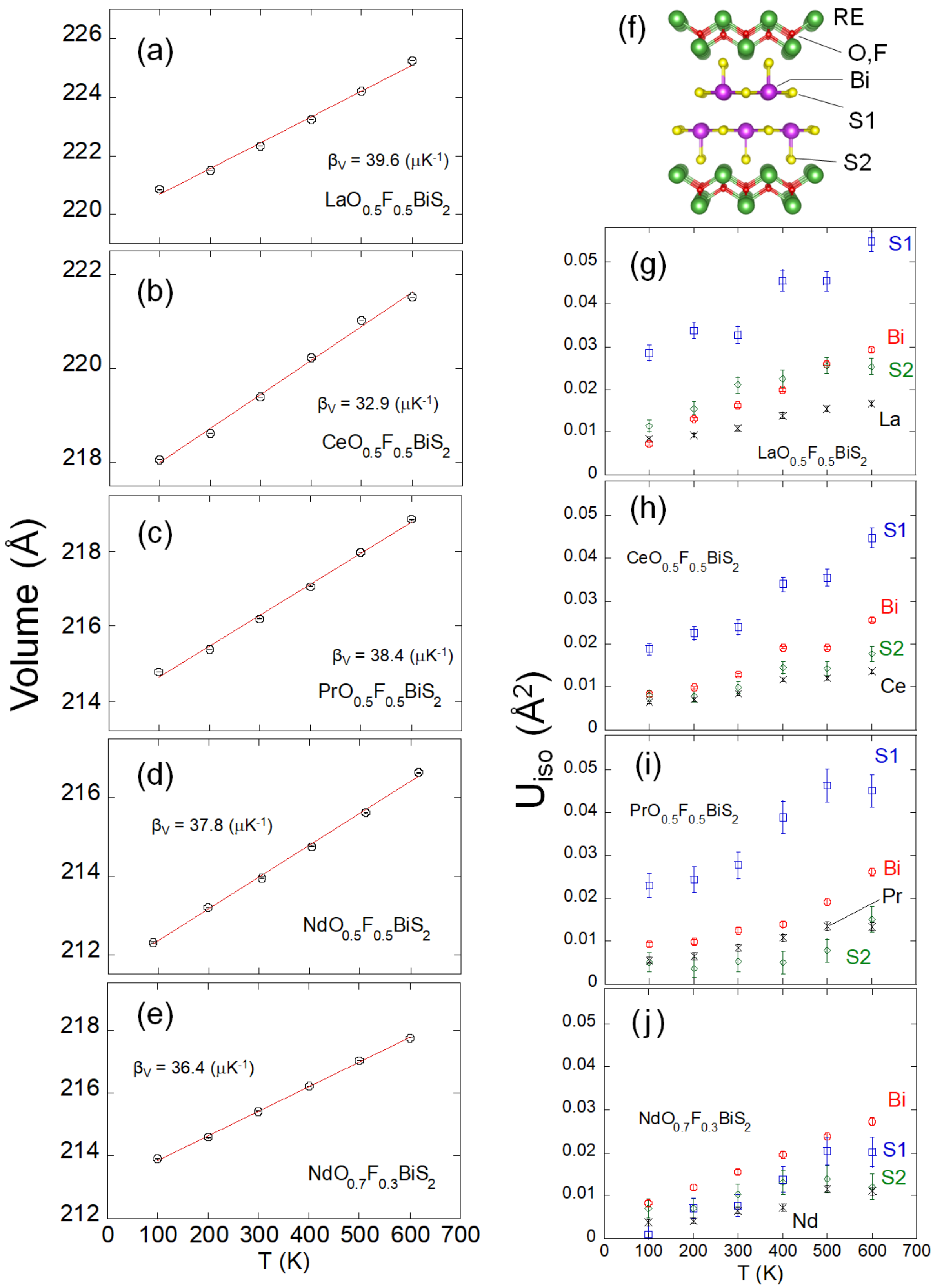

Fig. 2. Temperature dependence of lattice volume ($V$) for (a) $LaO_{0.5}F_{0.5}BiS_2$, (b) $CeO_{0.5}F_{0.5}BiS_2$, (c) $PrO_{0.5}F_{0.5}BiS_2$, (d) $NdO_{0.5}F_{0.5}BiS_2$, and (e) $NdO_{0.7}F_{0.3}BiS_2$. (f) Schematic image of crystal structure of RE(O,F)$BiS_2$. Temperature dependence of isotropic atomic displacement ($U_{iso}$) for (g) $LaO_{0.5}F_{0.5}BiS_2$, (h) $CeO_{0.5}F_{0.5}BiS_2$, (i) $PrO_{0.5}F_{0.5}BiS_2$, and (j) $NdO_{0.7}F_{0.3}BiS_2$.

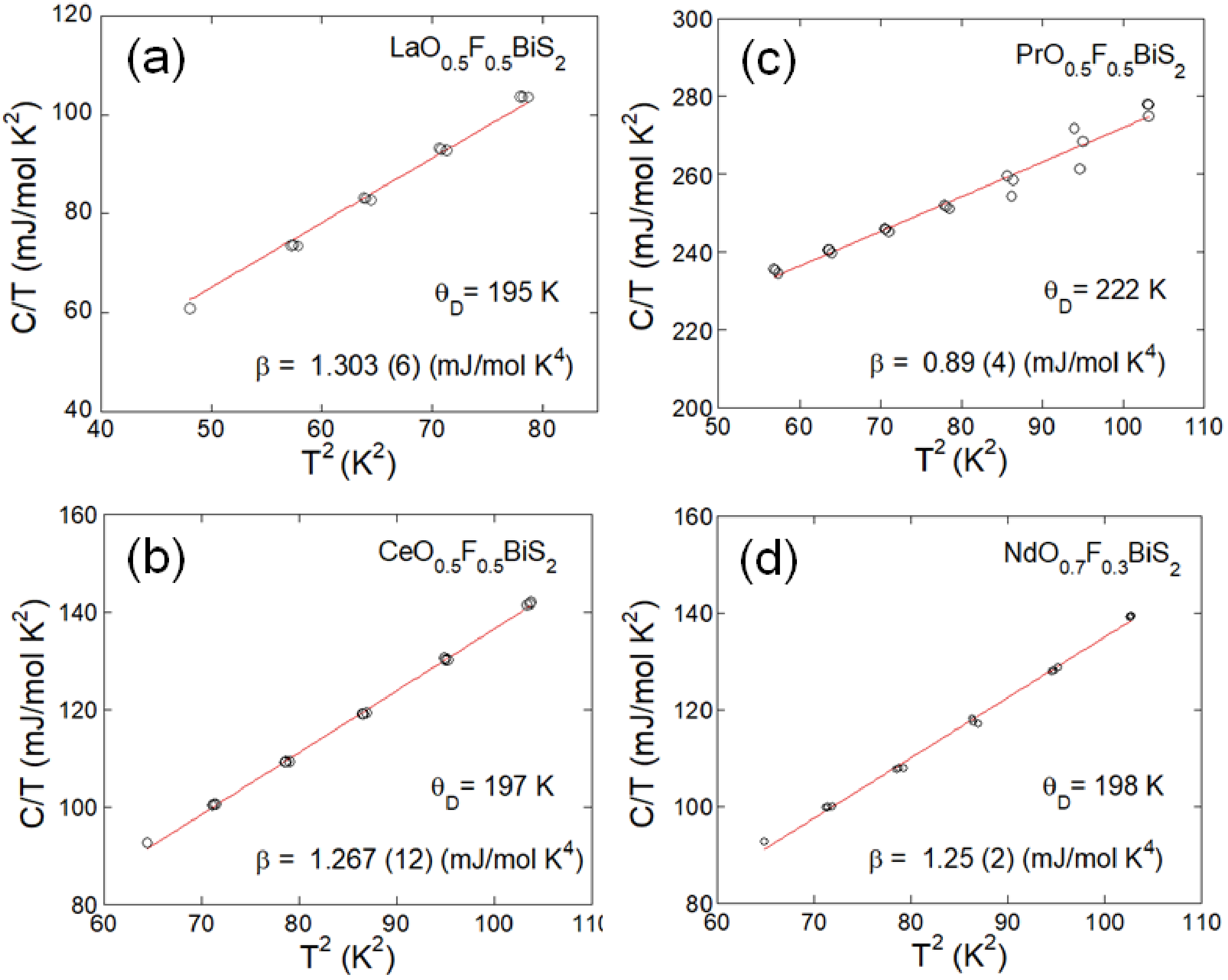

Fig. 3. $T^2$ dependence of $C/T$ for (a) $LaO_{0.5}F_{0.5}BiS_2$, (b) $CeO_{0.5}F_{0.5}BiS_2$, (c) $PrO_{0.5}F_{0.5}BiS_2$, and (d) $NdO_{0.7}F_{0.3}BiS_2$.

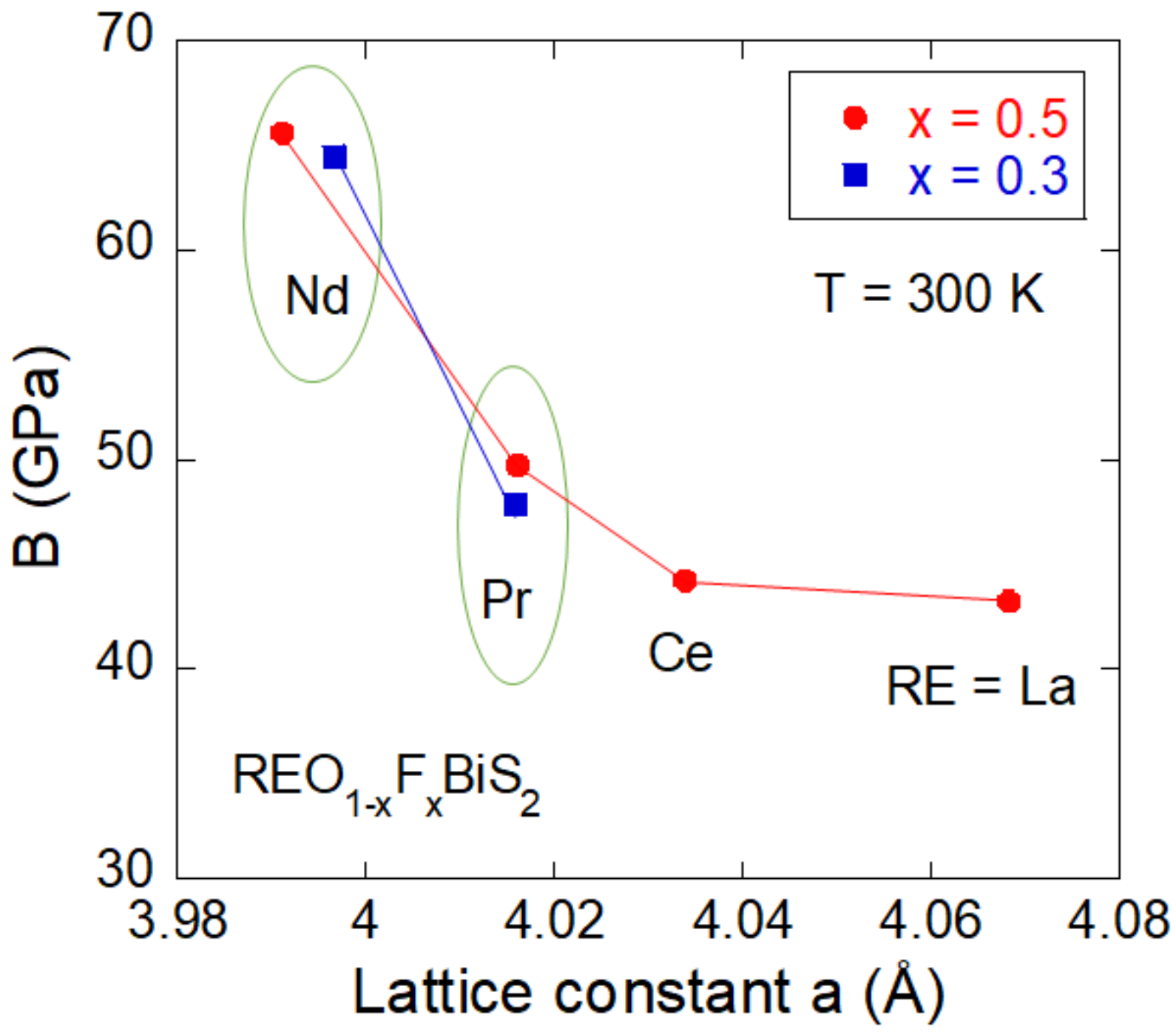

Fig. 4. Lattice constant ($T$ = 300 K) dependence of bulk modulus ($B$). For $REO_{1-x}F_xBiS_2$, lattice constant a becomes a good measure of in-plane chemical pressure. $B$ for $PrO_{0.7}F_{0.3}BiS_2$ has been reported in Ref. 50.

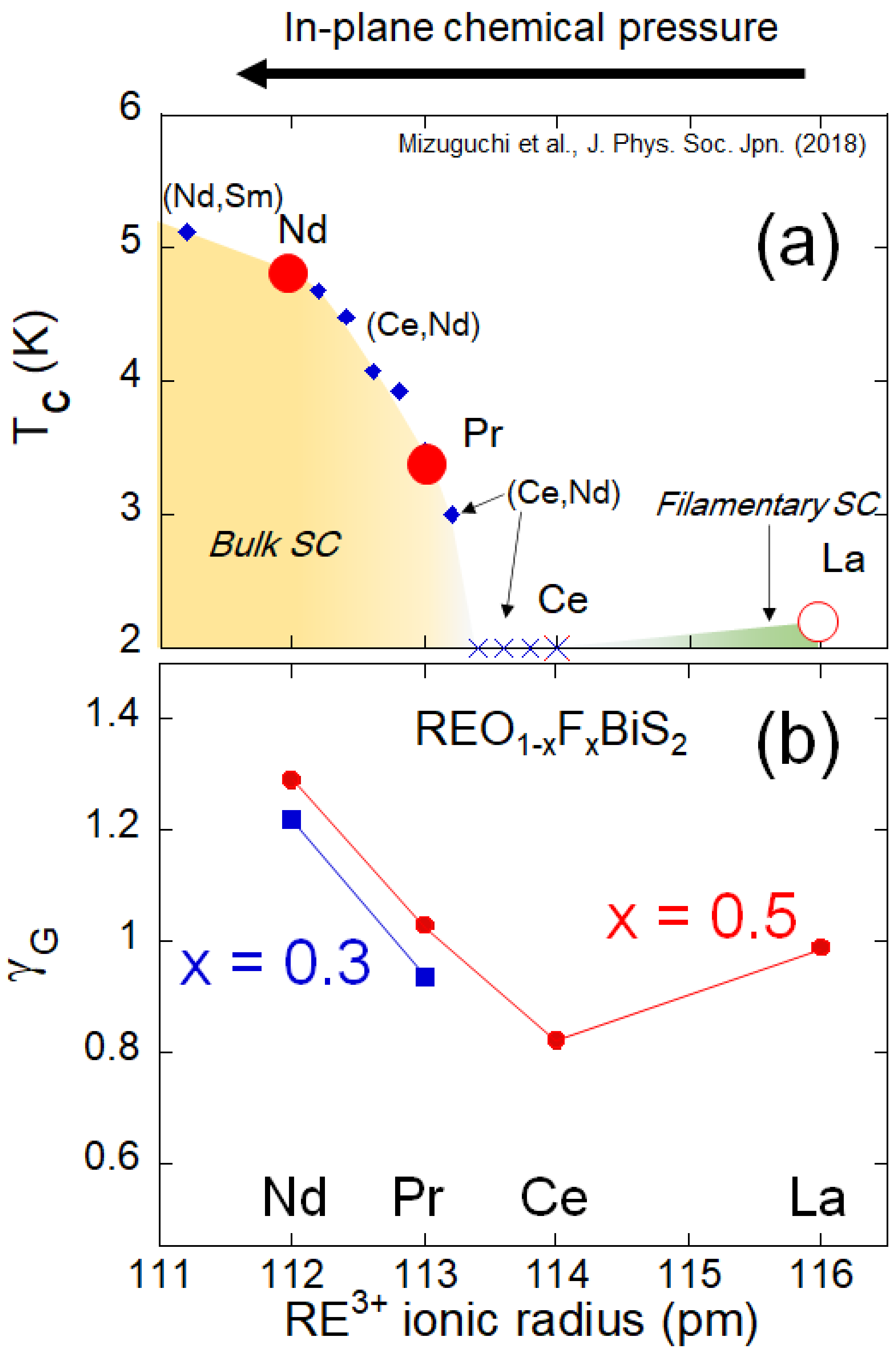

Fig. 5. $RE^{3+}$ ionic radius dependence of (a) $T_c$ for $REO_{0.5}F_{0.5}BiS_2$ ($x$ = 0.5) (reproduced from Ref. 26) and (b) estimated $\gamma_G$ for $REO_{1-x}F_xBiS_2$. In (a), cross symbols indicate that the compositions do not exhibit superconductivity at above 1.8 K.